\documentclass[a4paper,twocolumn,11pt,accepted=2024-12-06]{quantumarticle}
\pdfoutput=1
\usepackage[utf8]{inputenc}
\usepackage[english]{babel}
\usepackage[T1]{fontenc}
\usepackage[numbers]{natbib}
\usepackage[english]{babel}
\usepackage{graphicx}
\usepackage{dcolumn}
\usepackage{bm}

\usepackage{amsmath, braket} 
\usepackage{amssymb}
\usepackage{xcolor}
\usepackage{bbold}
\usepackage{hyperref}
\usepackage{comment}
\usepackage{tikz}  
\usetikzlibrary{arrows,shapes,positioning,shadows,svg.path,backgrounds,fit}

\newcommand{\op}[1]{\hat{#1}}
\newcommand{\tcb}[1]{\textcolor{black}{#1}}

\newcommand{\tcr}[1]{\textcolor{black}{#1}}
\newcommand{\tcrg}[1]{\textcolor{black}{#1}}
\newcommand{\tcbg}[1]{\textcolor{black}{#1}}
\begin{document}

\title{Negative Wigner function by decaying interaction from equilibrium}

\author{Michal Kolář}
\email{kolar@optics.upol.cz}
\affiliation{Department of Optics, Palack\'y University, 17. listopadu 12, 771 46 Olomouc, Czech Republic}
\author{Radim Filip}
\email{filip@optics.upol.cz}
\affiliation{Department of Optics, Palack\'y University, 17. listopadu 12, 771 46 Olomouc, Czech Republic}
\begin{abstract}
Bosonic systems with negative Wigner function superposition states are fundamentally witnessing nonlinear quantum dynamics beyond linearized systems and, recently, have become essential resources of quantum technology with many applications. Typically, they appear due to sophisticated combination of external drives, nonlinear control, measurements or strong nonlinear dissipation of subsystems to an environment. Here, we propose a conceptually different and more autonomous way to obtain such states, avoiding these ingredients, using purely sudden interaction decay in the paradigmatic interacting qubit-oscillator system weakly coupled to bath at thermal equilibrium in a low-temperature limit. We demonstrate simultaneously detectable unconditional negative Wigner function and quantum coherence and their qualitative enhancement employing more qubits.   
\end{abstract}

\maketitle

\section{Introduction}
Modern quantum physics and technology depend on nontrivial quantum superpositions \cite{StreltsovRevModPhys2017,ChitambarRevModPhys2019,degenRevModPhys2017}. They are typically driven by a classical, coherent, strong external force that builds required quantum coherence \cite{baumgratz2014}. At the next level, it is advantageous when, instead of an external drive, the coherent energy source is encapsulated in the interaction with the thermal bath \cite{giacomoPRL2018} or, even better, in the free system Hamiltonian \cite{kolar2023local}. Then, no external coherent strong drive is needed and quantum resources appear more autonomously. Such cases require synthetic processes that unconditionally create coherences within an individual system from the thermal energy population and redistribute them where needed. Until now, all the experiments have been proposed for earning and accumulating quantum coherence of the single two-level system (qubit) without requiring a coherent measurement to induce coherence {\color{black}\cite{giacomoPRL2018,RicardoPRA2021,slobodeniuk2021extraction,kolar2023local}}. 
 
Linear quantum oscillators are opposite cases to the qubits saturable at the first excited level. If coherent external force linearly drives the oscillator, Gaussian coherent states arise \cite{glauberPhysRev1963}. Respective of a phase reference, a shot-noise-limited laser can reach such classical coherences from thermal energy \cite{glauber2PhysRev1963,glauberRevModPhys2006}. Classical coherences are still compatible with classical coherence theory widely used for energy 
absorptive measurement \cite{Mandel_Wolf_1995}. Combined with energy non-conserving linearized oscillator dynamics, nonclassical coherences can rise by coherent, strong parametric drives in Gaussian squeezed coherent states \cite{yuenPhysRevA1976,yuenPhysRevLett1983,cavesPhysRevD1983}. Although 
these coherences can improve quantum sensing \cite{degenRevModPhys2017} and quantum communication \cite{khanRSTA2017,LaviePhysRevApplied2022}, they are still simple for fundamental investigation and insufficient in advanced 
bosonic applications in quantum simulations, computing \cite{AspuruGuzikRevModPhys2020} and thermodynamics \cite{Lostaglio2015,Narasimhachar2015}. Such bosonic applications need quantum non-
Gaussian coherences beyond a convex set of the Gaussian ones. \tcrg{The bosonic applications use the advantage of the superposition of higher Fock states in the single bosonic mode. It contrasts different approaches based on spatial or other mode structures of individual bosons or classical states, which do not exploit quantum bosonic noise in the single mode \cite{PittmanPRA1995,AbouraddyPRA2001,HowellPRL2004,WALBORN201087,Aspden_2013,Douce2013,Hor-MeyllPRL2014,Moreau2018,Moreau2019,Ianzano2020,Ndagano2020}. The experimental implementation of nonlinearity in a single mode is essential to gain these bosonic advantages, and it cannot be substituted by mode shaping. } 

\tcr{Non-autonomous (externally driven and controlled) transient generation of quantum non-Gaussian states is covered by a considerable amount of literature} \tcrg{\cite{WalschaersPRXQuantum2021,CAI202150,LACHMAN2022100395,RAKHUBOVSKY2024100495}}. \tcrg{On the other hand, only a few methods focus on a non-autonomous generation of the quantum non-Gaussian states in steady state \cite{JoanaPRA2016,QuijandrPRL2018,QuijandrPRR2021,Rips2012}. 
To the best of our knowledge, no method discusses the more autonomous steady-state generation of non-Gaussian states of a linear oscillator without an external drive, as is possible for two-level systems \cite{kolar2023local}.}

Our preliminary investigation has shown that a straightforward extension of the qubits' methodology \cite{kolar2023local} is insufficient if a coherent measurement is not used. Using thermal equilibrium states as initial states of the entire system, without any external coherent drive, only mixtures of Gaussian quantum coherence can be conclusively generated.
By coherently measuring the subsystems, one can observe conditional steering of quantum non-Gaussian coherence. However, finding an unconditional method of coherence generation without any other measurement requires new resources and approach to be used. 

Here, to reach quantum non-Gaussian coherence unconditionally from thermal equilibrium, we propose to exploit a sudden decay of a part of interactions in combination with synthetic qubit-oscillator interactions. We analyze this effect in detail and observe further upgrades of unconditional quantum non-Gaussian state generation using more qubits employing such decaying interactions with the oscillator. Such approach uncovers a hidden and unexploited potential of interaction decays to reach quantum non-Gaussian states without external coherent drives and measurements, and later use them in applications.

{\color{black} We first introduce our model in Sec.~\ref{sec-system}. Section~\ref{sec-protocol} describes the protocol we propose for achieving non-Gaussian states with negative Wigner function. Next,  in Sec.~\ref{sec-interactions} we qualitatively describe the main working principles of the proposed protocol. In Section~\ref{sec-negativity} we introduce the functionals used to quantify negativity of the Wigner function. The non-Gaussian core states achievable in our protocol are characterized in Sec.~\ref{sec-coherence}. \tcb{The primary steps towards a fully autonomous \tcbg{protocol} are presented in Sec.~\ref{sec-autonomous}.} Finally, the roles of various parameters specifying our model are discussed in Sec.~\ref{sec-discussion}.
}

\begin{figure*}[ht]
\includegraphics[width=1.75\columnwidth]{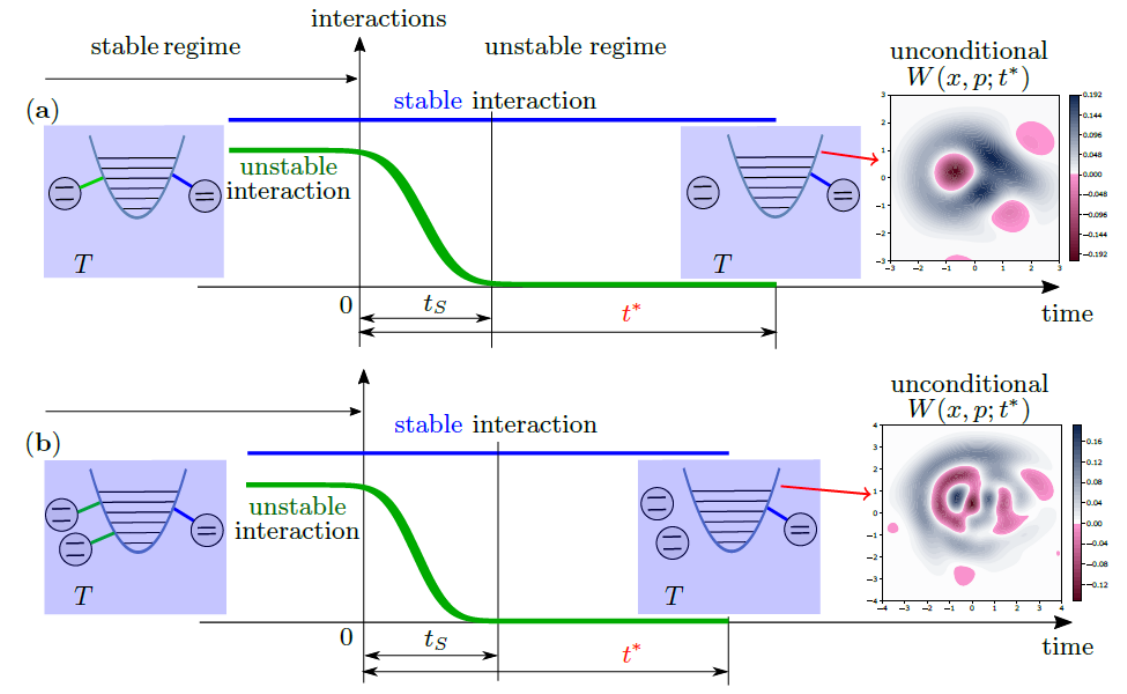}
\caption{Schematic of the protocol for the generation of negative Wigner function due to decaying interaction. (a) The linear harmonic oscillator (LHO) is interacting with one two-level system (TLS) via a {\it stable} transversal interaction (blue colored lines) and another ancillary TLS via {\it unstable} longitudinal interaction (green colored lines). Simultaneously, the total system is {\it continuously and weakly} interacting (blue rectangles) with a common thermal bath at temperature $T$. During the stable regime, this thermal bath prepares thermal state of the total system. At time $t=0$, the unstable ancillary interaction {\it decays} (green curves) on a (possibly variable) time scale $t_S$, while the stable interaction (blue curve) and weak interaction with the bath are constantly present. After suitable time $t^*$, the remaining stable interaction generates LHO reduced state with negativities (pink regions) of the Wigner function, although the interaction with the bath is still present. (b) The protocol employing {\it a pair} of ancillary TLSs interacting with the LHO. Quantitative effect of such pair is larger compared to the case of a single ancillary TLS, allowing to obtain more complex regions of negativities in (possibly) shorter evolution times. }
\label{fig1}
\end{figure*}

\section{Decaying longitudinal interaction}
\label{sec-system}
Conceptually, we consider a special compound system (an artificial molecule) consisting of several mutually coupled subsystems with different stability of the couplings, see Fig.~\ref{fig1}. Differently to previous work \cite{kolar2023local}, our target inside such a molecule is a linear harmonic oscillator (LHO) coupled to two (or more) intra-molecular two-level systems (TLSs). Each of these TLSs is separately coupled to the LHO via interactions with a different stability.  

Such molecule, within the desired coupling regimes, can be modeled by the respective Hamiltonians (setting $\hbar =1$) as \cite{wang2023NatComm}
\begin{eqnarray}\nonumber
\op{H}_1&=&\omega\,\op{n}+\frac{\omega_0}{2}\op{\sigma}_z+\frac{\omega_A}{2}\op{\sigma}_z^A+g_{A}(t)\,\op{\sigma}_z^A\op{X}+g_R \op{\sigma}_x\op{X},\\
\op{H}_2&=&\omega\,\op{n}+\frac{\omega_0}{2}\op{\sigma}_z+\frac{\omega_A}{2}\left(\op{\sigma}_z^{A_1}+\op{\sigma}_z^{A_2}\right)\label{eq-H-Rabi-zxdriven-direct}\\
\nonumber
&&+g_{A}(t)\,\left(\op{\sigma}_z^{A_1}+\op{\sigma}_z^{A_2}\right)\op{X}+g_R \op{\sigma}_x\op{X},\\
\nonumber
&&g_A(t)=g_A,\quad t\leq 0,
\end{eqnarray}
in the case of a single ancillary TLS, or of a pair of ancillary TLSs, respectively, which will serve the purpose of a quantitative comparison of the studied effects. In Eq.~\eqref{eq-H-Rabi-zxdriven-direct}, $\omega$ is the LHO frequency \cite{GU20171,Liu_2014,PhysRevA.91.053820,PhysRevLett.98.257003,PhysRevB.91.094517,PhysRevLett.115.203601}, $\omega_0$ the TLS frequency, $\omega_A$ stands for the frequency of the ancillary TLS(s), $g_A(t)$ for the (unstable) decaying longitudinal coupling strength between ancillary TLS(s) and LHO, and $g_R$ represents the stable strength of the transversal interaction (here of the Rabi type) between LHO and another TLS. The deep-strong coupling regime $g_{A(R)}/\omega\approx 1$ \cite{wang2023NatComm} considered below is a prerequisite to observe the effects. \tcb{The time dependent coupling $g_A(t)$, as given implicitly in Eq.~\eqref{eq-H-Rabi-zxdriven-direct}, does not in principle differentiate between {\it autonomous} case with classical degree of freedom evolving through certain dynamical equations of motion yielding in turn the values of $g_A(t)$, and  classical external {\it driving}, i.e. non-autonomous case. More detailed description of such interesting  autonomous case is elaborated in Sec.~\ref{sec-autonomous} and Appendix~\ref{sec:s-autonomous}, while the non-autonomous version sufficient, e.g., for a proof-of-principle experimental realization of our protocol, is being discussed in Sec.~\ref{sec-discussion} and Appendix~\ref{sec:s-decay}.    }

\section{The protocol}
\label{sec-protocol}
The complete protocol assumed throughout this paper is sketched in 
Fig.~\ref{fig1}. We emphasize that the protocol does not assume any external drive or strong nonlinear coupling to an external environment. The approach we suggest relies on the assumption, that the total system described by the Hamiltonians $\op{H}_{1(2)}$, Eq.~\eqref{eq-H-Rabi-zxdriven-direct}, is initially prepared in a global, {\it low-temperature} thermal state $\op{\tau}$ ($k_B=1$ in the following)
\begin{eqnarray}
\op{\tau}&=&Z^{-1}\exp\left(-\frac{\op{H}}{T} \right),\,Z={\rm Tr} \left[\exp\left(-\frac{\op{H}}{T} \right)\right],
\label{eq-thermal-tau}
\end{eqnarray}
due to the assumed presence of a thermal bath {\it weakly} coupled to the total system, with $\op{H}$ representing $\op{H}_{1(2)}$. Having $\op{\tau}$, Eq.~\eqref{eq-thermal-tau}, as the initial state, at time $t=0$ the interaction coefficient $g_A(t)$, Eq.~\eqref{eq-H-Rabi-zxdriven-direct}, starts to suddenly decay (vanish) on a certain timescale $t_S$, i.e., $|g_A(t_S)|\rightarrow 0$, see green curve in Fig.~\ref{fig1}, due to an assumed intrinsic instability of this interaction. While $\op{\tau}$ represents a steady state of the time evolution for {\it fixed} Hamiltonian parameters and thermal bath with temperature $T$ a swift change, relative to typical timescales in $\op{H}$ and thermalization with the bath, in the value $g_A(t)$ turns it into a non-stationary state with respect to the same thermal bath, thus initiating dynamics of the total system. Such evolution is governed by the stable part of (here the blue curve in Fig.~\ref{fig1}) interaction between the TLS and LHO and by the appropriate global master 
equation capturing the weak coupling of the {\it total} system and bath. At this point, we emphasize that exchanging the stability properties of the respective interaction coefficients $g_{A(R)}$, cannot generate any of the desired effects described below. 

\begin{figure*}[ht]
\includegraphics[width=1.95\columnwidth]{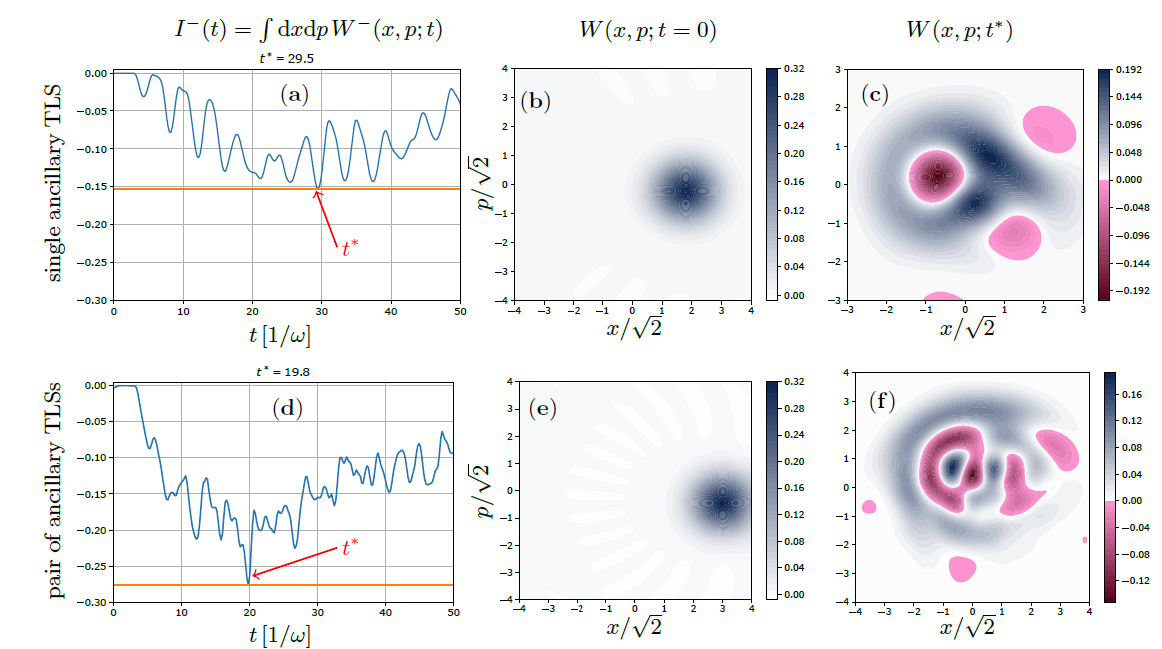}
\caption{Quantitative results (negativity of Wigner function and quantum coherence) of the protocols described in Fig.~\ref{fig1}. The first row of panels (a--c) shows the results for a single ancillary TLS coupled to LHO, see {\color{black}Eq.~\eqref{eq-H-Rabi-zxdriven-direct}}, whereas the second row (d--f) shows an example of results for protocol employing a pair of ancillary TLS interacting with LHO. The first column of panels represents the time dependent integrated Wigner function negativity $I^-(t)$, Eq.~\eqref{eq-neg-int}, quantifying non-classicality of the LHO reduced state. The second column of panels shows the initial Wigner functions of LHO at $t=0$. The last column presents the snapshots of Wigner functions with the optimal (the deepest) negativity $I^-(t^*)$ (pink regions) at suitable time $t^*$, which is shorter in the case of a pair of ancillary TLS. The resulting negativities of respective $W(x,p;t^*)$ span larger region of the phase space in the second case. In both cases shown, decoupling of the ancillary TLS(s) happens at $t=0$, when the total system is in the thermal state {$\op{\tau}$, \color{black} Eq.~\eqref{eq-thermal-tau},} and is instantaneous, $t_S\rightarrow 0$. Subsequent dynamics of LHO is governed by the Rabi and bath interactions. In all cases the parameters are $\omega_0=2$, $\omega=1$, $\omega_A=2.4$, $g_R=0.6$, $g_{A}(0)=0.8$, $T=2\cdot 10^{-2}$, and $\kappa=2\cdot 10^{-3}$, see {\color{black}Eq.~\eqref{eq-H-Rabi-zxdriven-direct}}. }
\label{fig2}
\end{figure*}

The master equation for the total system density matrix $\op{\rho}$ can be formally written in the Bloch-Redfield master equation (BRME) form \cite{JOHANSSONqutip2013}
\begin{eqnarray}
\partial_t\op{\rho}=-i\left[\op{H}(t),\op{\rho}\right]+\kappa\mathcal{R}(t)\left[\op{\rho}\right],
\label{eq-BRME}
\end{eqnarray}
with $\op{H}(t)$ the Hamiltonian~\eqref{eq-H-Rabi-zxdriven-direct}, coefficient $\kappa\ll\omega$ determines the weak coupling to the bath, $\mathcal{R}(t)$ the Bloch-Redfield tensor (superoperator), and we have explicitly indicated their possible time-dependence. Due to the infinite dimension and relative complexity of our system~\eqref{eq-H-Rabi-zxdriven-direct}, we resort ourselves to numerical solution of the system dynamics allowing to capture all necessary information about the system. Specifically, we use an approach based on well-established numerical implementation in QuTiP \cite{JOHANSSONqutip2013}, allowing as well for solving open system dynamics 
with time-dependent parameters {\color{black}\cite{PhysRevLett.131.143602,AnnPhysRevLett2023,Iyama2024NatComm}}. 
The typical example of the evolution is shown in Fig.~\ref{fig2} for certain suitable choice of the system parameters. We stress that these chosen values need not to be ``fine tuned'' in any sense, as presented below, meaning the quality of the discussed effect being stable with respect to the parameters' choice.

As the global system evolves according to Eq.~\eqref{eq-BRME}, we focus on the properties of the {\it reduced} LHO state, namely on its quantum coherence in local energy eigensbasis and non-Gaussianity represented by negative values of the corresponding Wigner function~\cite{Wigner}, at certain optimal time of evolution denoted as $t^*$ in this work, with the corresponding state denoted as $\op{\rho}^*$.

\section{The working principle}
\label{sec-interactions}
\begin{figure}[ht]
\includegraphics[width=.95\columnwidth]{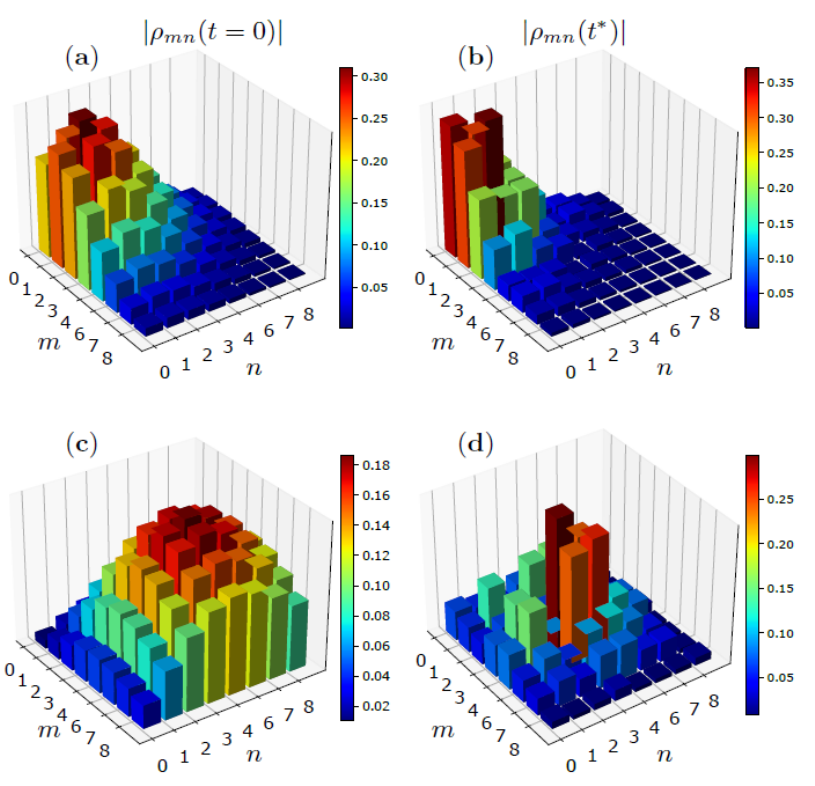}
\caption{{\color{black}Results for the total LHO state coherence in the Fock basis. The first row reflects single ancillary TLS interaction case, whereas the second row applies to a pair of ancillary TLSs. In the first column, we plot the modules of the reduced LHO state at $t=0$ for the first eight Fock states. The second column shows the final, Wigner function negativity-optimized, LHO reduced state at $t=t^*$, see Fig.~\ref{fig2}. In all cases the parameters are $\omega_0=2$, $\omega=1$, $\omega_A=2.4$, $g_R=0.6$, $g_{A}(0)=0.8$, $T=10^{-2}$, and $\kappa=10^{-3}$, see Eq.~\eqref{eq-H-Rabi-zxdriven-direct}.} }
\label{fig3}
\end{figure}

The protocol introduced in the previous section assumes the presence of two qualitatively different interactions, having two distinct purposes. The first interaction, named here ancillary and/or unstable, corresponds to the Hamiltonian~\eqref{eq-H-Rabi-zxdriven-direct} term $g_A(t)\,\op{\sigma}_z^A\op{X}$. As mentioned in the previous Sec.~\ref{sec-protocol}, the 
explicit time-dependence of $g_A(t)$ reflects our assumption that this interaction effectively decays on a time scale $t_S$, after a long-enough 
stable regime, see Fig.~\ref{fig1} green line, preceding this decay and allowing for initial thermalization of the whole artificial molecule modeled by~\eqref{eq-H-Rabi-zxdriven-direct}. 
\tcb{Such explicit time dependence may effectively result, e.g., from the presence of other degrees 
of freedom or from active switching-off by some external agent. This second option is clearly non-autonomous, as mentioned in Sec.~\ref{sec-system}, already, and could simplify \tcbg{proof-of-principle 
experimental tests of the interaction mechanism}. The preferred and truly autonomous option relies on extension of our system by other (possibly classical) degree of freedom whose evolution will match the decay of $g_A(t)$ depicted, e.g., in Fig.~\ref{fig1}. Such ``build, release, and hands-off'' approach is described in more detail in Sec.~\ref{sec-autonomous}.} The purpose of this interaction is to {\it autonomously} prepare $\op{\tau}$, thermal state~\eqref{eq-thermal-tau} with suitable 
properties. Namely, unstable interaction induces a nonzero initial displacement of the LHO state, see Fig.~\ref{fig2}(b), with Gaussian Wigner function despite the global thermalization. Hence, the corresponding LHO density matrix $\op{\rho}$, see Fig.~\ref{fig3}(a), has certain dominant populations of Fock states $\ket{n},\,n\geq 1$ (in that particular case $n=1,2$) and corresponding nonzero off-diagonal terms, indicating certain coherence between these Fock states. 

The second interaction, denoted here as stable, is in our model  Hamiltonian~\eqref{eq-H-Rabi-zxdriven-direct} represented by the term $g_R\,\op{\sigma}_x\op{X}$. It is assumed to be stable during the whole evolution, see Fig.~\ref{fig1} blue line. Our choice of the interaction form represents the Rabi interaction~\cite{RabiPhysRev1937,LvPRX2018} between LHO and TLS. Its purpose is to transfer excitations between these subsystems during quantum evolution. 

The working principle of our protocol can be sketched as follows. Decay of the unstable interaction initializes the subsequent evolution of the LHO+TLS subsystem, see Eq.~\eqref{eq-BRME} and Fig.~\ref{fig2}, which can be qualitatively described in the following {\it simplified} consideration based on Jaynes-Cummings model (JCM)~\cite{EberlyPRL1980}. Neglecting for the sake of simplicity the effect of the bath presence and assuming resonance of LHO and TLS, let the subsequent unitary evolution be initialized in a product of the weak coherent state and ground state of the stably coupled TLS. This stable interaction tends to transfers the populations of LHO Fock states to the TLS. In the first part of the evolution TLS is partially excited and subsequently this excitation is transferred back in the ongoing evolution. During such transfer, each of the participating Fock states exchanges the population with TLS on a different timescale, similarly as in JCM~\cite{EberlyPRL1980}. In the backward (TLS to LHO) population transfer, on a timescale corresponding to re-population of the (initially) dominantly populated Fock states, the lower and higher Fock states are still and again, respectively, captured in the TLS population or interaction. On a relevant timescale, this effectively  keeps the coherent contributions of the dominant Fock states after the LHO re-population and partially filters-out the remaining Fock contributions, similarly to collapses and revivals in JCM \cite{EberlyPRL1980}, but for low average photon numbers in our case.

Along with the LHO initial displacement, the unstable interaction $g_A(t\leq 0)\neq 0$ induces weak {\it coherence} of the stable TLS described approximately by the state $\ket{\psi}=\sqrt{p}\ket{e}+\sqrt{1-p}\ket{g}$, as a second order effect mediated by the stable interaction. Assuming again a simplified JCM type of evolution initialized in a product state of TLS state $\ket{\psi}$ and vacuum of LHO, after a proper evolution time, this is  swapped into a coherent superposition of Fock states $\ket{0}$ and $\ket{1}$, thus showing negativity of its Wigner function with estimated value 
\begin{eqnarray}
W_{\rm min}^{{\rm TLS}}\lesssim -\frac{p}{\pi}\exp\left[-\frac{1-p}{2p} \right],
\label{eq-neg-estimate}
\end{eqnarray}
where $p$ determines the {\it initial} probability of excitation of TLS. 

These two simplified qualitative considerations both result into a LHO reduced state characterized by negativity of its Wigner function. Our full model, with evolution governed by Eq.~\eqref{eq-BRME}, is much richer, as it takes simultaneously into account both coherent effects described above, the mutual LHO-TLS correlations stemming from the initial thermalization~\eqref{eq-thermal-tau} of the total system, and permanent presence of thermal bath during the whole evolution, Fig.~\ref{fig1}, even in the case of time-dependent Hamiltonian.

\section{Negativity of wigner function }
\label{sec-negativity}
We will quantify the Wigner function negativity in two different ways, yielding complementary information about the Wigner function structure. The first possibility is the integrated negativity $I^-(t)$ defined as \cite{AnatoleKenfack2004}
\begin{eqnarray}\label{eq-neg-int}
I^-(t)&=&\int {\rm d}x{\rm d}p\,W^-(x,p;t),\\ 
\nonumber
W^-(x,p;t)&=&{\rm max}\left[-W(x,p;t),0\right].
\end{eqnarray}

The second definition we adopt, is the global minimum (over the quadratures $x$, $p$) $N(t)$ of the Wigner function defined as
\begin{eqnarray}
N(t)={\rm min}_{(x,p)}W(x,p;t).
\label{eq-neg-min}
\end{eqnarray}
By the above mentioned complementarity of these quantities we mean that the integral definition of $I^-(t)$ is complete, but requires full state tomography, whereas the local definition of $N(t)$ is simpler, directly measurable negativity witness \cite{Chabaud2021witnessingwigner,LuPRL2021negativity}.

Figure~\ref{fig2} shows the main results of our analysis in terms of $I^-(t)$, Eq.~\eqref{eq-neg-int}. Namely, panels (c) and (f) reveal signatures of non-Gaussian coherence (as well as negativity $N$) of LHO, appearing at proper evolution time denoted $t^*$.  The negativities of Wigner function are represented by the pink regions in  contour plots. The effect of generating LHO states with negative Wigner function in the proposed protocol is quite stable with respect to the range of values of various parameters in the model, meaning that the working point can be chosen at will, bearing in 
mind several loose qualitative restrictions discussed in Sec.~\ref{sec-discussion}. 

\begin{figure}[ht]
\includegraphics[width=.95\columnwidth]{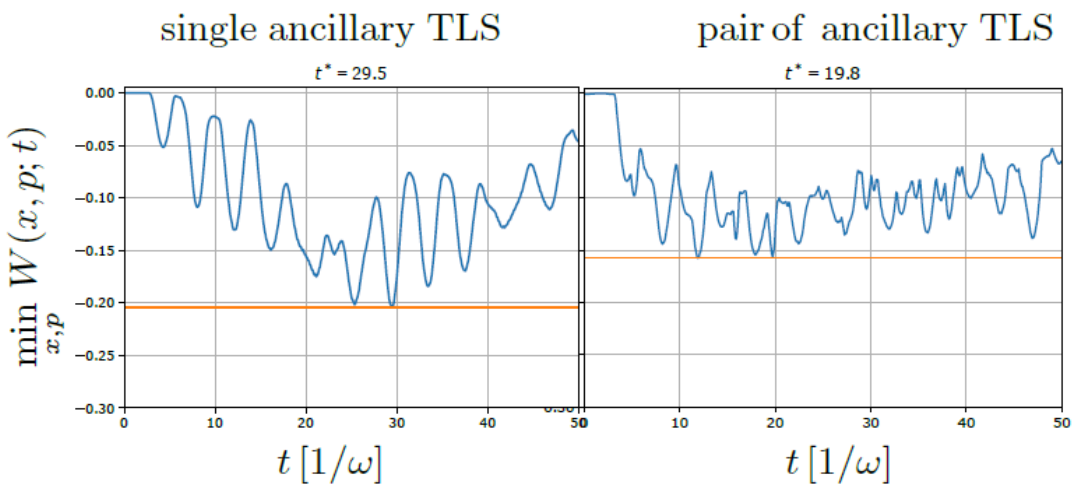}
\caption{ Global minimum of the Wigner function versus time of evolution in different settings as in Fig.~\ref{fig2} and parameters used there. For a single ancillary TLS (left panel) the global minimum corresponds to $\op{\rho}(t^*)$ from Fig.~\ref{fig3}(b) and so does the Wigner function in Fig.~\ref{fig2}(c). }
\label{fig-neg-mins}
\end{figure}
\section{Non-gaussian coherence}
\label{sec-coherence}

To reveal more precisely the non-Gaussian (NG) character of the negativity optimized LHO state $\op{\rho}_i(t^*)$, where indices $i=1,2$ label the single or a pair of ancillary TLSs case, respectively, we can conveniently apply general Gaussian operations $\op{G}$ (displacement and squeezing) to the state $\op{\rho}_i^*\equiv\op{\rho}_i(t^*)$ and characterize its non-Gaussian core state \cite{MenziesPRA2009} $\op{\rho}_i^{NG}=\op{G}_{\rm opt}\op{\rho}_i^*\op{G}^\dag_{\rm opt}$ by {\it minimizing} the Shannon entropy $S(\op{G}\op{\rho}_i^*\op{G}^\dag)$ of the Fock states occupation probabilities. We would like to emphasize that we do not intend to discuss and/or derive any NG criteria \cite{LachmanPRL2019} in this work. 

Let us point out again that such operations have to be applied numerically due to the complexity of our system. 
These operations include general displacement $\op{D}$ and squeezing $\op{S}$ of the state $\op{\rho}_i^*$ 
\begin{equation}
\op{\rho}_i(s,\beta)=\op{S}(s)\op{D}(\beta)\op{\rho}_i^*\op{D}^\dag(\beta)\op{S}^\dag(s),
\label{eq-NG-optimization}
\end{equation} 
parametrized by $\beta$ and $s$, respectively, over which the optimization is performed, to yield $\op{\rho}_i^{NG}\equiv \op{\rho}_i(s_i,\beta_i)$ with optimal values $\{s_i,\beta_i\}$. For actual numerical evaluation, we use as a working point the set of parameters employed in Fig.~\ref{fig2}, i.e., $\omega_0=2$, $\omega=1$, $\omega_A=2.4$, $g_R=0.6$, $g_{A}(0)=0.8$, $T=2\cdot 10^{-2}$, and $\kappa=2\cdot 10^{-3}$ and used throughout the paper as a typical reference point in the parameter space.

\begin{figure}[ht]
\includegraphics[width=.95\columnwidth]{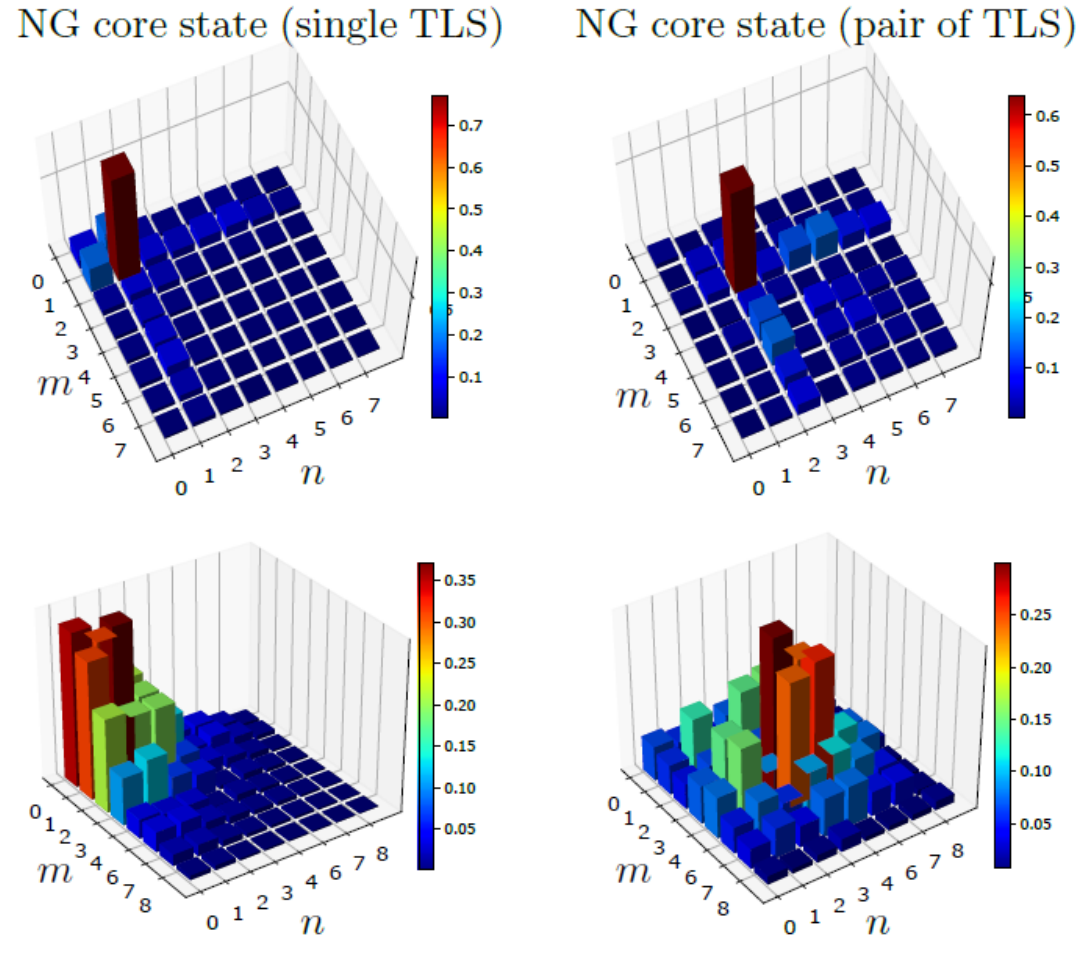}
\caption{{\color{black} Upper row: Modules $|\rho_{mn}|$ of the density matrix elements of  optimal core state obtained by optimizing~\eqref{eq-NG-optimization}: (left) $\op{\rho}_1^{NG}$ for a single ancillary TLS and (right) $\op{\rho}_2^{NG}$ for a pair of ancillary TLS interacting with LHO, see Sec.~\ref{sec-coherence}.  For comparison, lower row presents the original optimal states $\op{\rho}^*$ resulting from the protocol, see Fig.~\ref{fig3}(b,d).} }
\label{fig-matrix-displ-sq-1i2A-rhoopt-comp}
\end{figure}

The results of these transformations for the above mentioned parameters yield in the case of a single ancillary TLS $\beta_1=0.44-i\;0.15$ and $s_1=0.13$, while displacement $\beta_2=0.48-i\;0.52$ and squeezing $s_2=-0.11$ is obtained in the case of a pair of TLSs. The resulting non-Gaussian core states are shown in Fig.~\ref{fig-matrix-displ-sq-1i2A-rhoopt-comp} upper row, for both settings. As one can notice in the single TLS case $\op{\rho}_1^{NG}$ is dominated by Fock state $\ket{1}$ with weak partially coherent contributions of $\ket{0}$ and $\ket{2}$.
In the case of $\op{\rho}_2^{NG}$ (a pair of TLSs) $\ket{2}$ dominantly contributes and again partially coherent contributions of $\ket{1}$ and $\ket{4}$ are present, see Fig.~\ref{fig-matrix-displ-sq-1i2A-rhoopt-comp} for their comparison with $\op{\rho}_i^*$.

\begin{figure}[ht]
\includegraphics[width=.95\columnwidth]{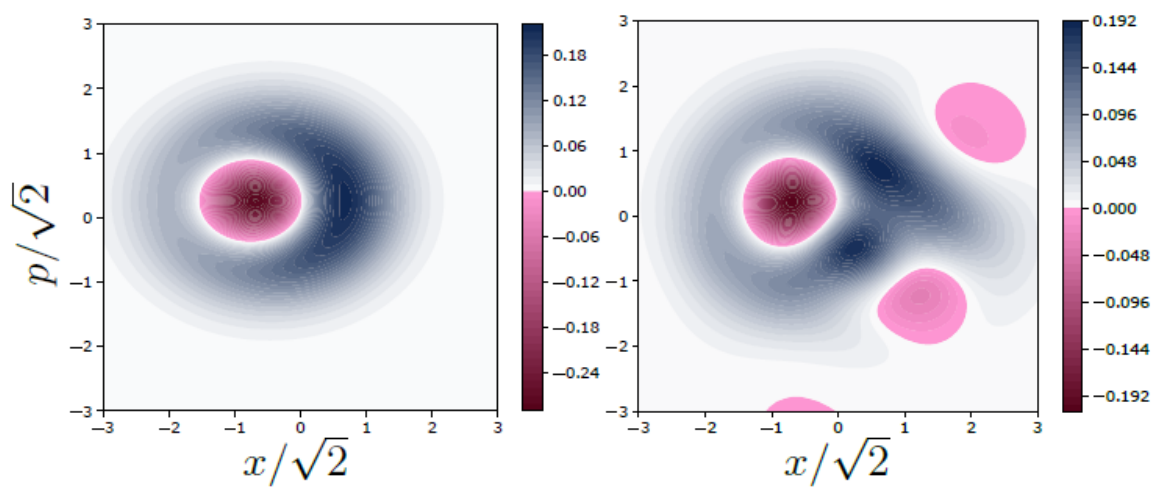}
\caption{ Comparison of the Wigner functions of fidelity-optimizing test state $\ket{\overline{\alpha},\overline{\xi},\overline{p}_0}$~\eqref{eq-test-state-def} (left), see Fig.~\ref{fig-fidelity3D} as well, and negativity-optimized state $\op{\rho}_1(t^*)$ (right) for a single ancillary TLS interacting with LHO.}
\label{fig-wigner-test-opt-1A}
\end{figure}

Complementary, we can approximately characterize the resulting negativity-optimized states $\op{\rho}_{1(2)}^{*}$ by a pure test state (``ansatz").  Such optimal test state can be obtained by maximization of the fidelity \cite{JozsaJMO1994}
\begin{eqnarray}
F(\rho,\sigma)=\left({\rm Tr}\sqrt{\sqrt{\rho}\sigma\sqrt{\rho}}\right)^2
\label{eq-fidelity-def}
\end{eqnarray}
of the negativity-optimized LHO density matrix $\rho^*_{1(2)}$, see Fig.~\ref{fig3}(b,d), with respect to $\alpha$-displaced and $\xi$-squeezed Fock states superposition. These superposition states are chosen based on the resulting non-Gaussian core states of the previous paragraph for each respective case. Namely, for the single ancillary TLS case the test state reads 
\begin{eqnarray}
\ket{\alpha,\xi,p_0}=\op{D}(\alpha)\op{S}(\xi)(\sqrt{p_0}\ket{0}+\sqrt{1-p_0}\ket{1}).
\label{eq-test-state-def}
\end{eqnarray}
Numerically found global maximum $\overline{F}[\rho^*_1,\ket{\alpha,\xi,p_0}\bra{\alpha,\xi,p_0}]=\bra{\alpha,\xi,p_0}\rho^*_1\ket{\alpha,\xi,p_0}$ is at $\overline{\alpha}=-0.34+i\,0.18$, $\overline{\xi}=-0.1$, and $\overline{p}_0=0.09$, yielding test-state fidelity $\overline{F}=0.84$, see Fig.~\ref{fig-wigner-test-opt-1A} for comparison of the corresponding Wigner functions. {\color{black} For this case, the global profile of the fidelity is shown in Fig.~\ref{fig-fidelity3D} for squeezing $\xi=\overline{\xi}$ set to 
the optimal value.}

\begin{figure}[ht]
\includegraphics[width=.95\columnwidth]{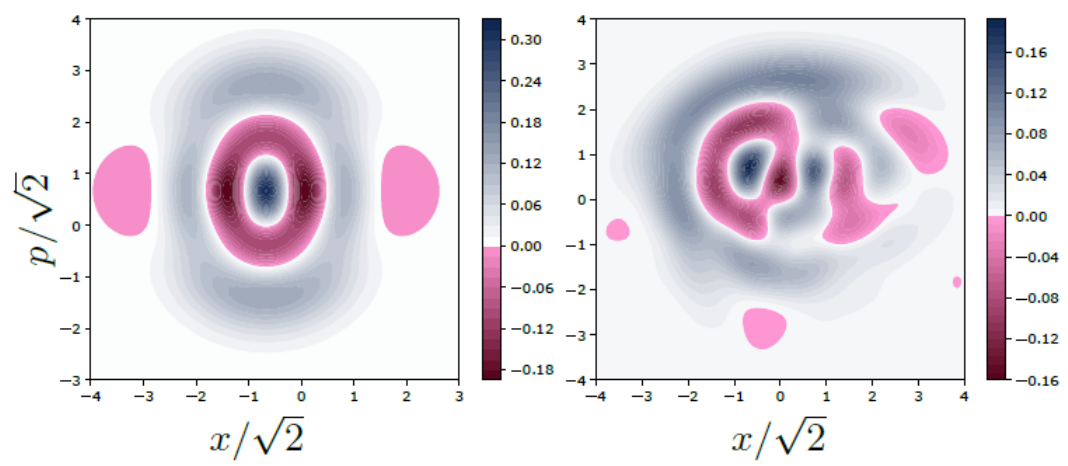}
\caption{The Wigner functions of fidelity-optimizing test state $\ket{\overline{\alpha},\overline{p}_2}$~\eqref{eq-test-state34-def} (left), and negativity-optimized state $\op{\rho}_2(t^*)$ (right) for a pair of ancillary TLSs interacting with LHO.}
\label{fig-wigner-test-opt-2A}
\end{figure}

In the second case of a pair of ancillary TLSs, the optimized test state reads
\begin{eqnarray}
\ket{\alpha,p_2}=\op{D}(\alpha)(\sqrt{p_2}\ket{2}+\sqrt{1-p_2}\ket{4}).
\label{eq-test-state34-def}
\end{eqnarray}
Numerically found global maximum $\overline{F}[\op{\rho}^*_2,\ket{\alpha,p_2}\bra{\alpha,p_2}]=\bra{\alpha,p_2}\op{\rho}^*_2\ket{\alpha,p_2}$ is found at $\overline{\alpha}=-0.47+i\,0.47$, $\overline{p}_2=0.87$, yielding optimal test-state fidelity $\overline{F}=0.67$, see Fig.~\ref{fig-wigner-test-opt-2A} for the corresponding Wigner functions. We have checked numerically that any squeezing $\xi\neq 0$ decreases the fidelity in this case. 
We describe just these two examples of the essential lowest-dimensional qubits and nontrivial superposition states in higher Fock states. After the first experimental tests of such primary cases, extensive exploitation with more TLS will be entirely possible. 
  
\section{Autonomous model}
\label{sec-autonomous}
\tcb{The autonomous evolution of a physical system governed by a Hamiltonian is usually understood as explicit time-independence of this Hamiltonian. On the other hand, if the Hamiltonian has some explicitly time-dependent parameter(s), it typically describes the case in which the system is coupled to some external {\it classical} agent actively steering those parameters, thus causing the time-dependence. The transition between these two situations might lead through extension of the system of interest by classical (or quantum) degree(s) of freedom (DOF) whose evolution implicitly represents the time dependence of the Hamiltonian parameters, e.g., governed by a suitable potential.}
 
\tcb{As we are primarily interested in {\it autonomous} generation of non-Gaussian states, we adopt the ``classical extension" option from the previous paragraph and assume that $g_A(t)$, Eq.~\eqref{eq-H-Rabi-zxdriven-direct}, effectively corresponds to the position $x(t)$ of a classical DOF which is evolving in a suitable potential $V(x)$. The resulting properties of $x(t)$ should match those of $g_A(t)$, i.e., $x(t)\equiv g_A(t)$, having long enough plateau allowing for thermalization of the global system towards $\op{\tau}$ \eqref{eq-thermal-tau} as governed by Eq.~\eqref{eq-BRME}, if the initial state represents vacuum and respective TLS's ground states. It is then followed by steep enough decay $|x(t)|\rightarrow 0$, allowing for sudden decoupling of the ancillary TLS, see the green curve in Fig.~\ref{fig1} as an example, leading to negativity $N(t)$ generation in the subsequent time interval, as in Sec.~\ref{sec-negativity}. As we discuss in the next Sec.~\ref{sec-discussion}, only these two qualitative properties are crucial for obtaining quantum non-Gaussian coherence in our protocol, unlike the precise decay profile. More technical details about the autonomous evolution can be found in Appendix~\ref{sec:s-autonomous}. }

\section{Discussion}
\label{sec-discussion}
By extensive numerical simulations of our system's time evolution, we conclude that in order to achieve the negativity of Wigner function, one has to employ strong coupling regime in both considered interaction constants, $g_{A(R)}\lesssim\omega$~\cite{wang2023NatComm}, as presented in 
Fig.~\ref{fig-negativity-gRgA}, see App.~\ref{s-coupling}. The effect is implied by an inseparable  presence of both initial state preparation employing $g_A(t\leq 0)\neq 0$ 
and possibility of the population transfer between LHO and TLS via $g_R\neq 0$. After reaching some threshold-value $g_A$ that decreases with increasing 
value of $g_R$, negativity improves almost linearly with $g_A$ within certain region of values. For $g_R\gtrsim 0.3$ (within the range of parameters we use in our example) the linear region is followed by local 
minimum with respect to $g_A$, revealing that further increase of $g_A$ will degrade the positive effect of achieving negativity. Thus, such achieved optimal value of negativity improves further along with 
increasing $g_R$, until reaching optimum at $g_R\approx 0.7$, whereas further increase of stable interaction constant again degrades negativity value. This can be attributed to the change of the ground state structure of $\op{H}$ in Eq.~\eqref{eq-H-Rabi-zxdriven-direct}, degrading NG character of optimized local LHO state.
Hence, around the working point used in our model, the optimum is approximately achieved by using values $g_R\approx 0.7,\,g_A(0)\approx 0.9$, yielding $N\approx -0.23$ for a single ancillary TLS case. In connection to these findings, we emphasize again that the (ultra)strong coupling effects can be effectively provided by using larger number of ancillary TLS with decaying interaction in the initial preparation stage of the protocol. Such settings provides larger initial displacements of LHO, hence larger coherent energy to begin with, and subsequently larger regions of the negativity quantified by $I^-(t^*)$ \eqref{eq-neg-int}, see Fig.~\ref{fig2}~(a,d), with not so deep negativity minima defined by $N(t^*)$ \eqref{eq-neg-min}, see Fig.~\ref{fig-neg-mins}. 

In this context short qualitative comparison with modified model in which Jaynes-Cummings model (JCM) like interaction between LHO and TLS is considered. Within the range of parameters used, JCM always performs worse in value of the achieved negativity showing positive impact of counter-rotating (CR) terms on negativity generation. A more detailed comparison of both models' results is beyond the scope of this paper as the results considerably depend on specific values of parameters used. One determining parameter is TLS frequency $\omega_0$, hence the LHO-TLS detuning $\Delta=\omega_0-\omega$. Not only $\Delta\neq 0$ does not forbid achieving negativity, but on contrary positive values of $\Delta >0$ deepen obtained negativity, see Fig.~\ref{fig-negativity-omega0}. Such fact can be attributed to the effect of CR terms allowing for population transfer even in the non-resonant $\Delta\neq 0$ conditions.

Another condition determining the achievability of NG states is necessity of maintaining low 
enough temperature $T$ defining the initial state $\op{\tau}$, Eq.~\eqref{eq-thermal-tau}. 
Our examples present the results from the vicinity of the working point whose parameters were specified, e.g. in the previous 
Sec.~\ref{sec-coherence}. They reveal that temperature values $T/\omega\approx 10^{-2}$ fully exploit the potential of the 
Hamiltonian~\eqref{eq-H-Rabi-zxdriven-direct} ground state to effectively achieve NG core states specified in Sec.~\ref{sec-coherence}. With increasing temperature $T$, the achieved minimum of negativity $N$ is less 
pronounced, as shown in Fig.~\ref{fig-negativity-T}. The increase of $T$ adds incoherently the contributions from excited 
states of $\op{H}_{1(2)}$, Eq.~\eqref{eq-H-Rabi-zxdriven-direct}, into the time evolution and increases thermal fluctuations, which turns out to be detrimental for the effect. Simultaneously, Fig.~\ref{fig-negativity-T} reveals existence of a plateau in 
region $T/\omega\lesssim 0.2$ for  values of parameters around our working point, offering acquiring almost the same negativity values in wider range of temperatures without the necessity of reaching $T\approx 0$ and confirming that the effect is not critically sensitive to temperature. 

Last parameter we examined in connection to NG states generation, is the type of decay-profile $g_A(t)$ and the corresponding typical  
timescale denoted here $t_S$, see Fig.~\ref{fig1}. We have explicitly tested two distinct decay profiles depicted schematically as the green curve on the schematic in Fig.~\ref{fig1}(a,b), namely Gaussian and  exponential one, see Eqs.~\eqref{eq-decay-profiles}. Both profiles collapse into instantaneous disappearance of the unstable interaction in the limit of $t_S\rightarrow 0$ and thus fully reproduce results presented in such limit throughout this paper, e.g. in Fig.~\ref{fig2}. We assume the profiles being submitted to a specific constraint, namely that they effectively vanish (decouple TLS from LHO) at the same time $t\approx 2t_S$, at which both reach $2\%$ of their initial value. In the intermediate regime $0\leq\omega t_S\lesssim 2$, the decrease of negativity values is  relatively slow, see Figs.~\ref{fig-negativity-tS},~\ref{fig-negativity-tS-exp-gauss}~(left column), while for $\omega t_S\gtrsim 3$ the negativity loss is more rapid, thus its final values, although time-optimized, become practically negligible, see Fig.~\ref{fig-negativity-tS-exp-gauss}~(right column). In this regime of slower decay, we recognize certain advantage of exponential decay-profile in terms of achievable negativity. This fact can be probably attributed to smaller time-integrated area under the exponential decay profile. In general, we can unify these observations by concluding that the shorter the decay time scale $t_S$ the more pronounced are the non-classical properties of the LHO final state.

\section{Outlook} 
{\color{black} As a stimulating future goal one can anticipate effort to gain more complex superposition of the Fock states in the core states of the evolution outcomes. As a preliminary check suggests, such goal can be in principle analyzed, again, in a fully numerical approach, aiming to a different figure of merit than the optimal negativity. Such core superposition-state complexity can be foreseen to increase with different duration of the decay-following LHO evolution. Additionally, the Fock state indices can be to a certain extent affected by increasing number of ancillary TLS. Although such simulations might require more computation time, further investigation of more autonomous quantum non-Gaussian state generation is desirable.

Here, we would also like to mention potential application in which NG states provide advantage over the classical states. Let us consider quantum-phase-estimation problem in which we employ the NG state $\op{\rho}_1^*$ (outcome of the protocol, e.g. Sec.~\ref{sec-protocol} and 
Fig.~\ref{fig2}(c)) and compare its performance to the one of coherent state $\op{\rho}^C$ with the same displacement, here $\{ \langle\op{X}\rangle _{\op{\rho}},\langle\op{P}\rangle _{\op{\rho}} \}$, as $\rho_1^{*}$ in the same quantum-phase $\theta$ estimation task. The performance of respective states in this task can be well quantified by the quantum Fisher information (QFI) \cite{BraunsteinPhysRevLett1994,MontenegroCommPhys2023}. The state in our numerical example can be characterized in this task  by ${\rm QFI}^{NG}(\op{\rho}_1^*,\theta)\doteq 3.69$, while ${\rm QFI}^{C}(\op{\rho}^C,\theta)\doteq 0.07$, both QFI quantities being independent of the phase $\theta$. The NG state shows a clear advantage over the corresponding Gaussian state similarly as in \cite{deng2023heisenberglimited}.  

In order to test our prediction experimentally, at least on the proof-of-principal level, the corresponding experimental platform has to allow for fulfillment of two conditions. Firstly, it has to allow for strong 
interactions of proper form between TLS(s) and LHO. Secondly, it has to have the ability of spontaneous thermalization of mutually interacting subsystems mentioned above with respect to their global energy eigenbasis. These demands might be jointly fulfilled in the case of superconducting circuits 
platforms \cite{ronzaniNature2018,PekolaRevModPhys2021,wang2023NatComm}, rendering them as a potentially suitable experimental platform to test our stimulating predictions. 

}

\section*{Acknowledgments}  {\color{black} The authors acknowledge support through Project No. 22-27431S of the Czech Science Foundation and project
CZ.02.01.01/00/22\_008/0004649 (QUEENTEC) of EU and MEYS Czech Republic.
The work was also supported by the European Union’s
2020 research and innovation programme (CSA
- Coordination and support action, H2020-
WIDESPREAD-2020-5) under grant agreement
No. 951737 (NONGAUSS). We also acknowledge
Horizon Europe Research and Innovation Actions
under Grant Agreement no. 101080173 (CLUSTEC).
}

\bibliography{refs}
\appendix

\section{The fidelity}
\label{sec:S-fidelity} 
\begin{figure}[ht]
\includegraphics[width=.9\columnwidth]{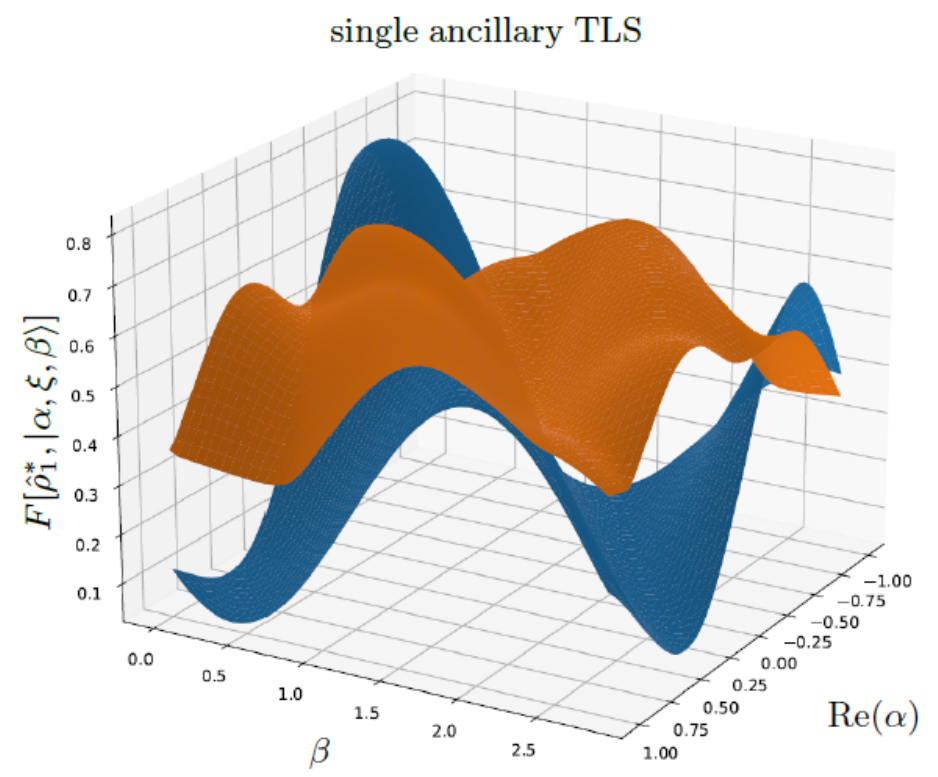}
\caption{Fidelity $F[\op{\rho}_1^*,\ket{\alpha,\xi,p_0}\bra{\alpha,\xi,p_0}]$, Eqs.~\eqref{eq-fidelity-def}--\eqref{eq-test-state-def}, of the negativity-optimized LHO density matrix $\op{\rho}_1^*$, see Fig.~\ref{fig2}, with respect to: (blue) $\alpha$-displaced, $\xi$-squeezed Fock states superposition $\ket{\alpha,\xi,\beta}=\op{D}(\alpha)\op{S}(\xi)(\sqrt{p_0}\ket{0}+\sqrt{1-p_0}\ket{1})$ with $p_0=\sin^2{\beta}$. Global maximum is at $\overline{\alpha}=-0.35+i\,0.18$, $\overline{\xi}=-0.11$, $\overline{p}_0=0.09$, yielding $\overline{F}=0.84$. (orange) Comparison of $\op{\rho}_1^*$ fidelity with {\it diagonal part} of the projector $\ket{\alpha,\xi,p_0}\bra{\alpha,\xi,p_0}$, yielding maximum at $\overline{\alpha}=0.35-i\,0.16$, $\overline{p}_0=0.1$, yielding $\overline{F}_D=0.6$, thus underlining importance of the test state's off-diagonal terms.}
\label{fig-fidelity3D}
\end{figure}

In this section we present a more global view showing, see blue landscape in Fig.~\ref{fig-fidelity3D}, dependence of the fidelity $F[\op{\rho}_1^*,\ket{\alpha,\xi,p_0}\bra{\alpha,\xi,p_0}]$, Eq.~\eqref{eq-fidelity-def}, of the negativity-optimized LHO density matrix $\op{\rho}_1^*$ (the case of a single ancillary TLS) on the displacement $\alpha$ and population $p_0=\sin^2\beta$ parametrizing state $\ket{\alpha,\xi,p_0}$, Eq.~\eqref{eq-test-state-def}. For comparison, we add as well the fidelity of state $\op{\rho}_1^*$ with fully dephased version of the test state $\ket{\alpha,\xi,\beta}$, yielding considerably lower value $\overline{F}_D=0.6$, thus underlining importance of the test state's off-diagonal terms, i.e. its coherence. The parameters used for the simulation are the same as in Fig.~\ref{fig2}.

For clarity reasons, we choose the fidelity to be plotted only in dependence on real part of the displacement parameter $\alpha$ and the mixing angle $\beta$, while the optimization was actually performed including imaginary part of $\alpha$ and squeezing parameter $\xi$. Thus, the plot is to be understood as a cut for ${\rm Im}(\alpha)=\{0.18,-0.16\}$, and $\xi=\{-0.11,0\}$ for the full test state, and its dephased version, respectively.

\begin{figure}[ht]
\includegraphics[width=.95\columnwidth]{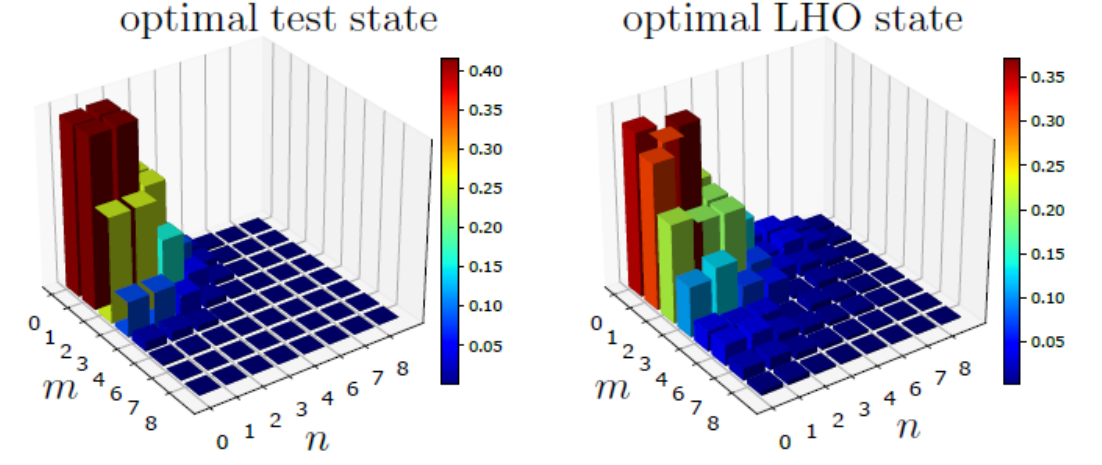}
\caption{Modules of the density matrix elements of (left) optimal test state $\ket{\overline{\alpha},\overline{\xi},\overline{p}_0}$, Eq.~\eqref{eq-test-state-def}, and (right) negativity-optimized state $\op{\rho}_1^*$, Fig.~\ref{fig3}(b), for a single ancillary TLS interacting with LHO. See Fig.~\ref{fig-fidelity3D} as well.}
\label{fig-matrix-test-opt-1A}
\end{figure}

For a more complete picture, the absolute values $|\rho_{mn}(t^*)|$ of the fidelity-optimized test state $\ket{\overline{\alpha},\overline{\xi},\overline{p}_0}$, Eq.~\eqref{eq-test-state-def}, are shown in Fig.~\ref{fig-matrix-test-opt-1A}~(left panel).

\section{The effects of coupling strengths}
\label{s-coupling}
\begin{figure}[ht]
\includegraphics[width=.9\columnwidth]{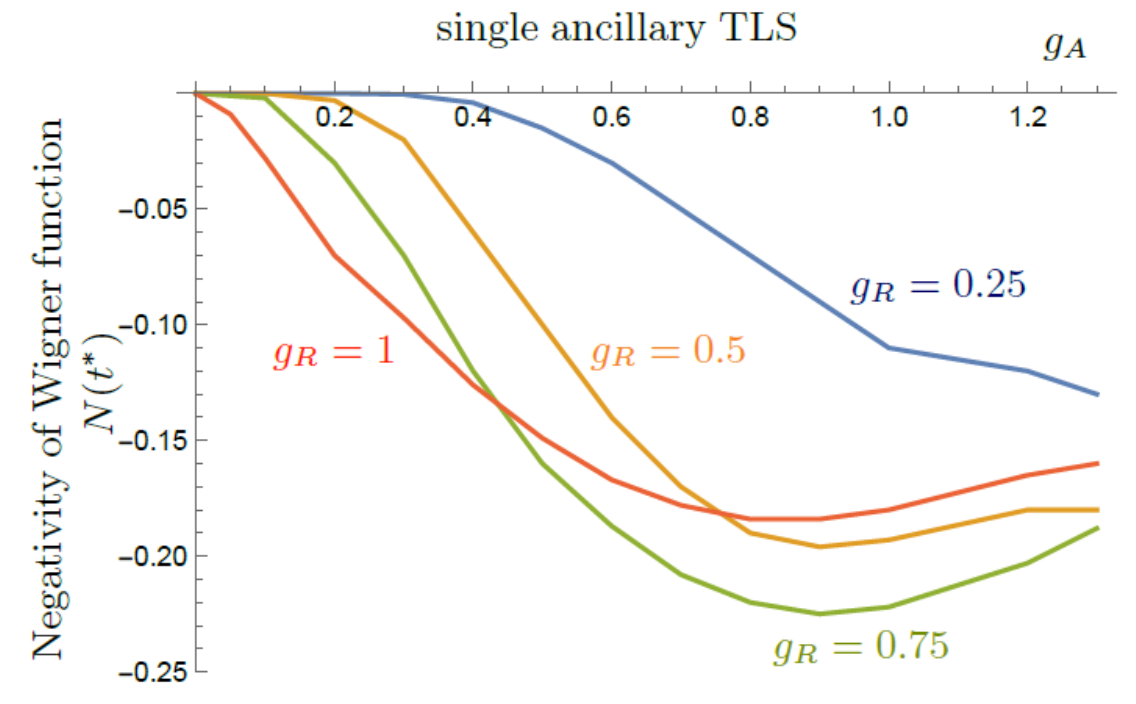}
\caption{Minimum negativity of the Wigner function $N(t^*)$ dependence on the initial value of unstable coupling constant $g_A$, with stable coupling $g_R$, see Fig.~\ref{fig1} and Eq.~\eqref{eq-H-Rabi-zxdriven-direct}, used as a parameter. The parameters used in numerical simulation are $\omega_0=2$, $\omega=1$, $\omega_A=2.4$, $T=2\cdot 10^{-2}$, and $\kappa=2\cdot 10^{-3}$, and instantaneous decay $t_S\rightarrow 0$ was assumed.}
\label{fig-negativity-gRgA}
\end{figure}

In Figure~\ref{fig-negativity-gRgA}, we show dependence of the optimized minimum of Wigner function negativity on the coupling of LHO to a single ancillary TLS $g_A$, see Eq.~\eqref{eq-H-Rabi-zxdriven-direct}, assuming instantaneous  ($t_S\rightarrow 0$) decay of $g_A(t)$. The curves are parametrized by values of stable coupling $g_R$. The plotted values were obtained by means of tracking the Wigner function negativity during the time evolution of the system (according to protocol of Fig.~\ref{fig1}), while the corresponding global minimum was recorded. For more details, see Sec.~\ref{sec-discussion} of the main text.

\section{Influence of the atomic frequency $\omega_0$}
\label{sec:s-atomic-freq}
\begin{figure}[ht]
\includegraphics[width=.92\columnwidth]{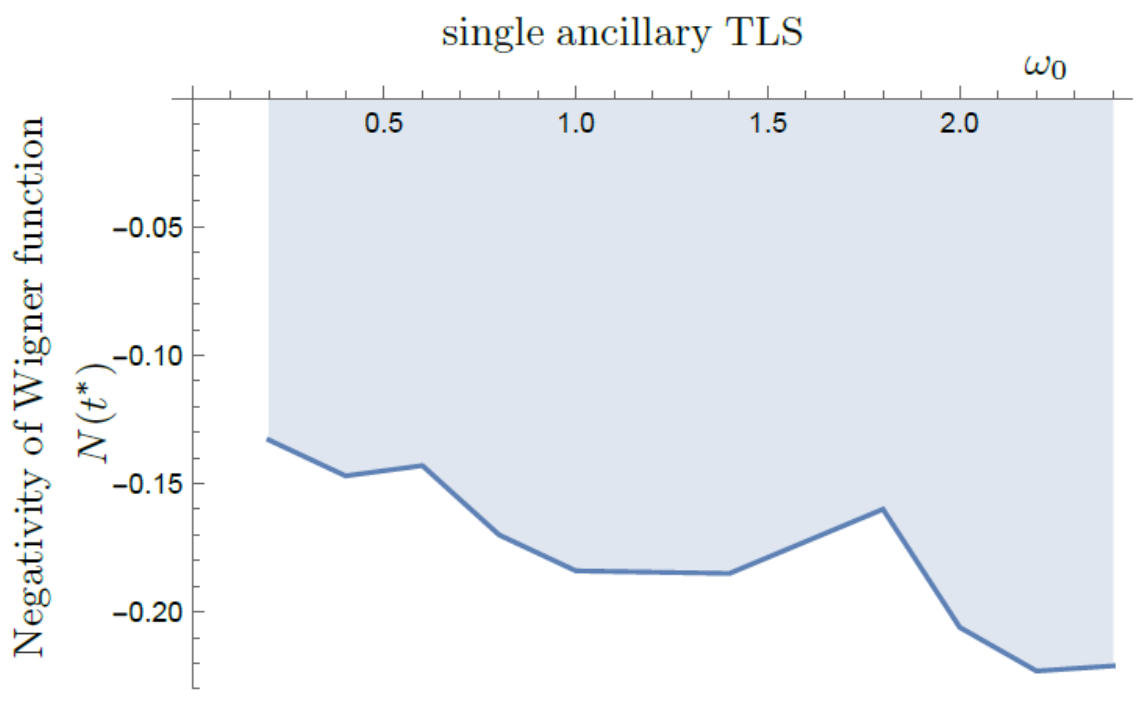}
\caption{Minimum negativity of the Wigner function $N(t^*)$~\eqref{eq-neg-min} dependence on the frequency $\omega_0$, see Eq.~\eqref{eq-H-Rabi-zxdriven-direct}. The parameters used in numerical simulation are $\omega=1$, $\omega_A=2.4$, $g_R=0.6$, $g_{A}(0)=0.8$, $T=2\cdot 10^{-2}$, and $\kappa=2\cdot 10^{-3}$, and instantaneous decay $t_S\rightarrow 0$ was assumed. }
\label{fig-negativity-omega0}
\end{figure}

Figure~\ref{fig-negativity-omega0} presents the optimized Wigner function negativity $N(t^*)$ dependence on atomic frequency $\omega_0$, see Eq.~\eqref{eq-H-Rabi-zxdriven-direct}. Other parameters are the same as in Fig.~\ref{fig2}. The obtained curve reveals a general trend of increasing the negativity with increasing atomic frequency $\omega_0$. The non-monotonic modulation (``toothy'' profile) is the outcome, see Fig.~\ref{fig-neg-mins}, of the beating of decreasing value of the slowly-changing envelope and fast on-top oscillations of the negativity values. The curves were obtained through tracking the Wigner function negativity during the time evolution of the system (according to protocol of Fig.~\ref{fig1}) and the corresponding global minimum was recorded in each run. For further information, see  Sec.~\ref{sec-discussion} of the main text.

\section{Temperature dependence}
\label{sec:S-temperature}
\begin{figure}[hbt]
\vspace{.5cm}
\includegraphics[width=.95\columnwidth]{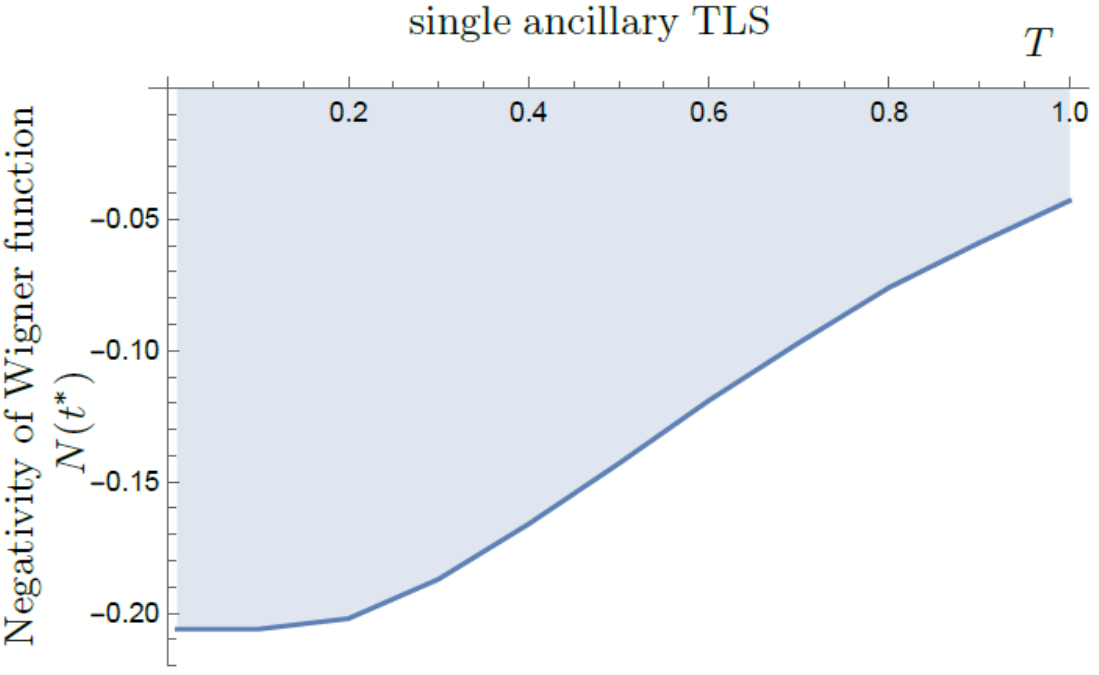}
\caption{Minimum negativity of the Wigner function $N(t^*)$ dependence on the temperature $T$ determining the thermal state $\op{\tau}$, see Eq.~\eqref{eq-thermal-tau} and App.~\ref{sec:S-temperature}. The parameters used in numerical simulation are $\omega_0=2$, $\omega=1$, $\omega_A=2.4$, $g_R=0.6$, $g_{A}(0)=0.8$, and $\kappa=2\cdot 10^{-3}$, and instantaneous decay $t_S\rightarrow 0$ was assumed.}
\label{fig-negativity-T}
\end{figure}

We plot the temperature $T$, see Eq.~\eqref{eq-thermal-tau}, dependence of the minimum of Wigner function negativity $N(t^*)$, Eq.~\eqref{eq-neg-min}, in Fig.~\ref{fig-negativity-T}. The plotted values were obtained based on the numerical time evolution for all parameters fixed, while in each such case the minimum of $N(t^*)$ was obtained and plotted and in the next step, the temperature $T$ was changed. For further information, see Sec.~\ref{sec-discussion} of the main text.

\section{Autonomous decay model}
\label{sec:s-autonomous}
\tcb{This Appendix aims to provide more technical details on the {\it autonomous} process of interaction $g_A(t)$ decay, Eq.~\eqref{eq-H-Rabi-zxdriven-direct}, under requirements outlined in Sec.~\ref{sec-autonomous}. These include considerable initial plateau, allowing for initial thermalization of the global system towards state $\op{\tau}$ \eqref{eq-thermal-tau} and subsequent sudden decay of $|g_A(t)|\rightarrow 0$, as depicted by the green line in Fig.~\ref{fig1}.} 

\begin{figure}[ht]
\begin{tikzpicture} 
 \node (img1)  {\includegraphics[width=.9\columnwidth]{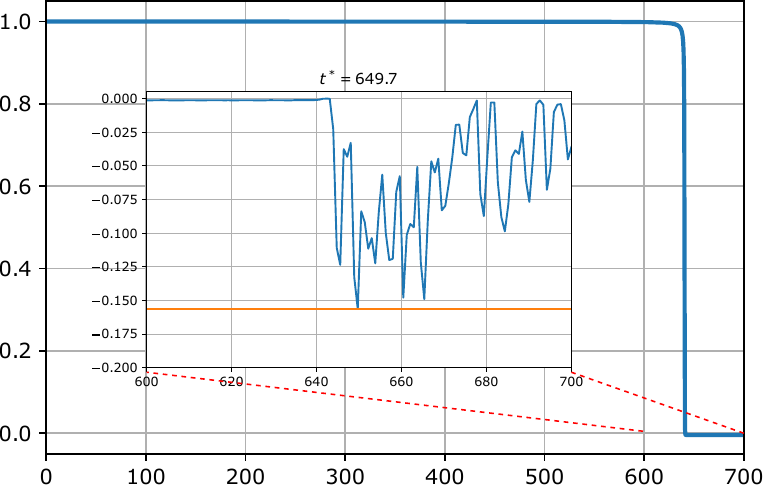}};
  \node[above=of img1, node distance=0cm, yshift=-6.6cm,xshift=.2cm] {{\color{black}$t\, [1/\omega]$}};
   \node[above=of img1, node distance=0cm, yshift=-1.cm,xshift=-0cm] {{\color{black}evolution\;of\;classical\;DOF\;and\;negativity}};
   \node[left=of img1, node distance=0cm, rotate=90, anchor=center, yshift=-.8cm,xshift=-0.cm] {{\color{black}$x(t)\equiv g_A(t)$}};
   \node[left=of img1, node distance=0cm, rotate=90, anchor=center, yshift=-1.9cm,xshift=0.1cm] {{\color{black}$N(t)$}};
\end{tikzpicture}
\caption{\tcb{Main panel: Numerical solution of Eq.~\eqref{eq-autonomous-diffeq} for $a=1.99$, $b=2.5$, $\epsilon =10^{-4}$, providing a plateau of length $t_p\approx 640$ for the chosen parameters. Subsequently, within interval $\Delta t\approx 4$ ancillary TLS and oscillator suddenly and autonomously decouple. Inset: This initializes Wigner negativity $N(t)$ generation, thus providing a fully autonomous version of our protocol starting from separable initial state of vacuum and respective ground states $\ket{0,g,g}$. The parameters' values used are: $\omega_0=1.5$, $\omega=1$, $\omega_A=2.5$, $g_R=0.6$, and $\kappa=9\cdot 10^{-3}$. } }
\label{fig-DOF-evolution}
\end{figure}
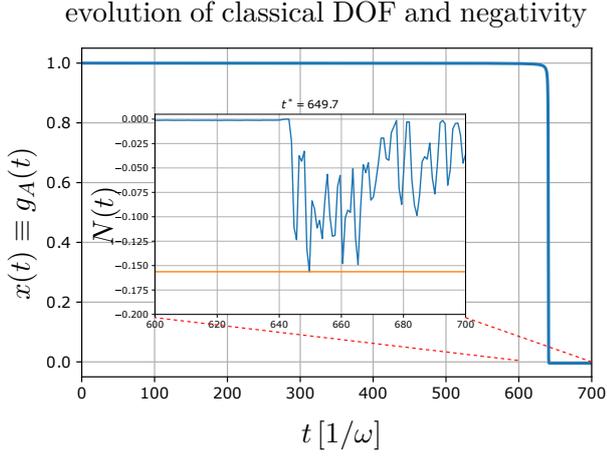

\tcb{To fulfill these, we assume that $g_A(t)$ is represented by the position $x(t)\equiv g_A(t)$ of additional degree of freedom (DOF) evolving in suitable \tcbg{nonlinear} potential $V(x)$. The profile of this potential is crucial to achieve the above mentioned properties of $x(t)$. We have chosen as \tcbg{an appropriate} candidate potential of the form}
\begin{eqnarray}
V(x)=\frac{a^4}{4}\left(1-x \right)^4-\frac{b^3}{3}\left(1-x \right)^3,\quad a,b>0.
\label{eq-autonomous-potential}
\end{eqnarray}
\tcb{As an important feature, this \tcbg{shifted cubic and quartic} potential has an inflection point at $x^*=1$, $V(x^*)=V'(x^*)=V''(x^*)=0$, providing a potential plateau in the vicinity this point.
\tcbg{Such locally unstable potentials are currently of experimental interest in the classical domain~\cite{OrnigottiPRE2018,OrnigottiPRL2018} and in future in the quantum domain~\cite{Rakhubovsky2021,Neumeier2024}, as well.}
We assume further that such degree of freedom evolves within that potential in an overdamped regime. By setting suitable initial condition $x(0)$, its equation of motion reads
\begin{eqnarray}
\frac{{\rm d}x}{{\rm d}t}=-\frac{{\rm d}V(x)}{{\rm d}x},\, x(0)=1-\epsilon,\, \epsilon \gtrsim 0.
\label{eq-autonomous-diffeq}
\end{eqnarray}
\tcbg{The construction of the potential \eqref{eq-autonomous-potential} together with setting the initial condition \eqref{eq-autonomous-diffeq} is to be considered as the {\it build} phase (of our protocol) in which the molecule, defined by Eq.~\eqref{eq-H-Rabi-zxdriven-direct}, is prepared. The evolution starts from \eqref{eq-autonomous-diffeq} in the classical DOF, together with vacuum and ground states of respective TLS's for the quantum system. After finishing this phase, the {\it release and hands-off} phase takes over, in which the molecule is created, and thermalized. The nonlinear equation} \eqref{eq-autonomous-diffeq} can be solved 
numerically, yielding $x(t)$ (the main panel of Fig.~\ref{fig-DOF-evolution}). \tcbg{The non-autonomous setting of the initial condition at the potential plateau implies a similar plateau-like behavior of $x(t)\equiv g_A(t)$, i.e. slowly varying (almost constant) position/coupling.} Such plateau, \tcbg{representing a stable period of the interaction $g_A(t)$,} allows for the above mentioned (almost perfect) thermalization of the quantum state close enough to $\op{\tau}$~\eqref{eq-thermal-tau}, if one \tcbg{builds such a device from initial separable} vacuum state and respective ground states $\ket{0,g,g}$. For the parameters used in our simulation, the fidelity~\eqref{eq-fidelity-def} of the actual evolved (according to Eq.~\eqref{eq-BRME}) state $\op{\rho}_f$ with respect to global ground state was $F\gtrsim 0.96$.}

\tcb{\tcbg{After the stable period $x(t)\approx 1$, a subsequent evolution provides, on contrary, a sudden decay} of the coupling, effectively switching it off, see main panel of Fig.~\ref{fig-DOF-evolution} around $t\approx 640$.  Such autonomous drop of coupling $g_A(t)$ initializes the evolution of quantum system governed by Eq.~\eqref{eq-BRME} and, \tcbg{in the subsequent time interval $t\in [642,700]$}, generates non-Gaussian (NG) states. These yield the negative Wigner function of the oscillator, see inset of Fig.~\ref{fig-DOF-evolution} around $t^*\approx 650$, similarly as in Fig.~\ref{fig2}.}

\tcb{In a nutshell, autonomous decay \tcbg{of the quasi-stable} coupling $g_A(t)$ requires initial preparation of global non-equilibrium state, which spontaneously evolves towards global equilibrium state and generates NG states in the course of this evolution.}

\section{Decay timescale and profile}
\label{sec:s-decay}
\begin{figure}[ht]
\includegraphics[width=.95\columnwidth]{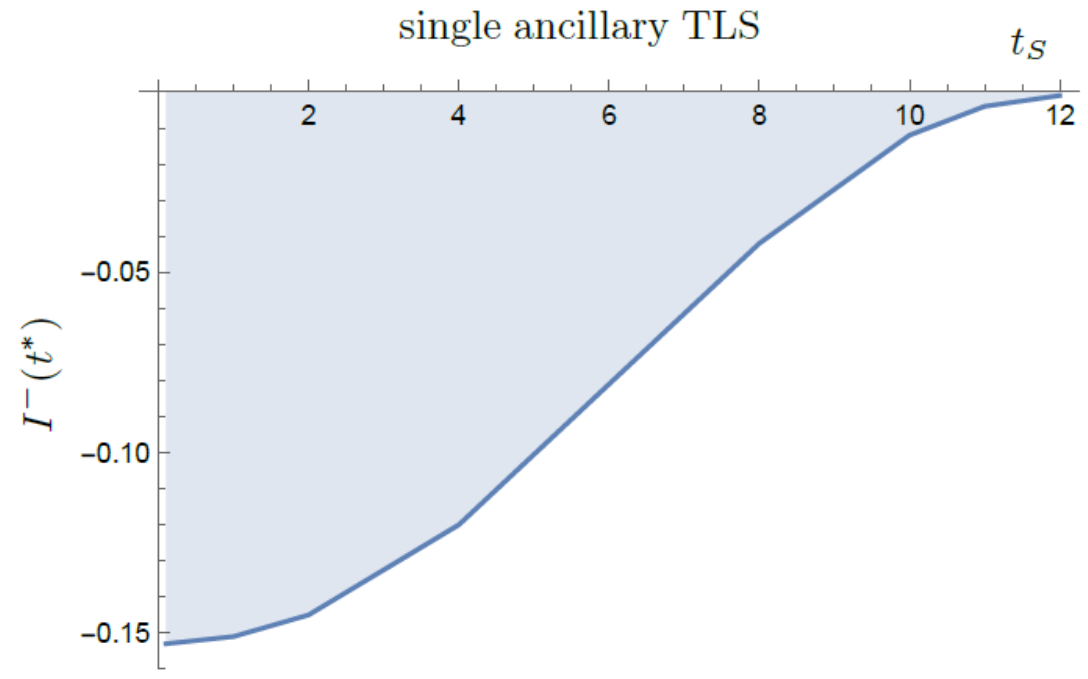}
\caption{Dependence of the integrated negativity $I^-(t^*)$, see Eq.~\eqref{eq-neg-int} on the decoupling timescale $t_S$. The plot was obtained in such way that for every fixed value of the decay time scale $t_S$ (and the rest of the parameters) the integrated negativity $I^-(t^*)$ for the Gaussian decay profile was tracked during the time evolution of the system (according to protocol of Fig.~\ref{fig1}) and the corresponding global minimum was recorded. The parameters used are the same as in Fig.~\ref{fig2}, i.e., $\omega_0=2$, $\omega=1$, $\omega_A=2.4$, $g_R=0.6$, $g_{A}(0)=0.8$, $T=2\cdot 10^{-2}$, and $\kappa=2\cdot 10^{-3}$.}
\label{fig-negativity-tS}
\end{figure}

In our numerical simulations, we were considering several decay profiles of the unstable interaction governed by $g_A(t)$. To allow for relevant quantitative comparison the decay profiles $g_A(t)$ were constrained by equality of their values at $t=0$ and $t=2\,t_S$ where we required that their values dropped to approximately $2\%$ of the initial value, hence being effectively switched-off. The particular profiles were chosen as
\begin{eqnarray}\nonumber
g_A(t)&=&\exp\left[-\left(\frac{t}{t_S}\right)^2 \right],\\
g_A(t)&=&\exp\left[-\frac{2\,t}{t_S}\right],
\label{eq-decay-profiles}
\end{eqnarray}
for Gaussian-like and exponential-like decays, respectively. The decoupling timescale $t_S$ considerably influences the achieved negativity of Wigner function in both cases. For the Gaussian-like profile, an example of such influence for the integrated negativity $I^-(t^*)$, see Eq.~\eqref{eq-neg-int}, is shown in Fig.~\ref{fig-negativity-tS}. 

\begin{figure}[ht]
\includegraphics[width=.95\columnwidth]{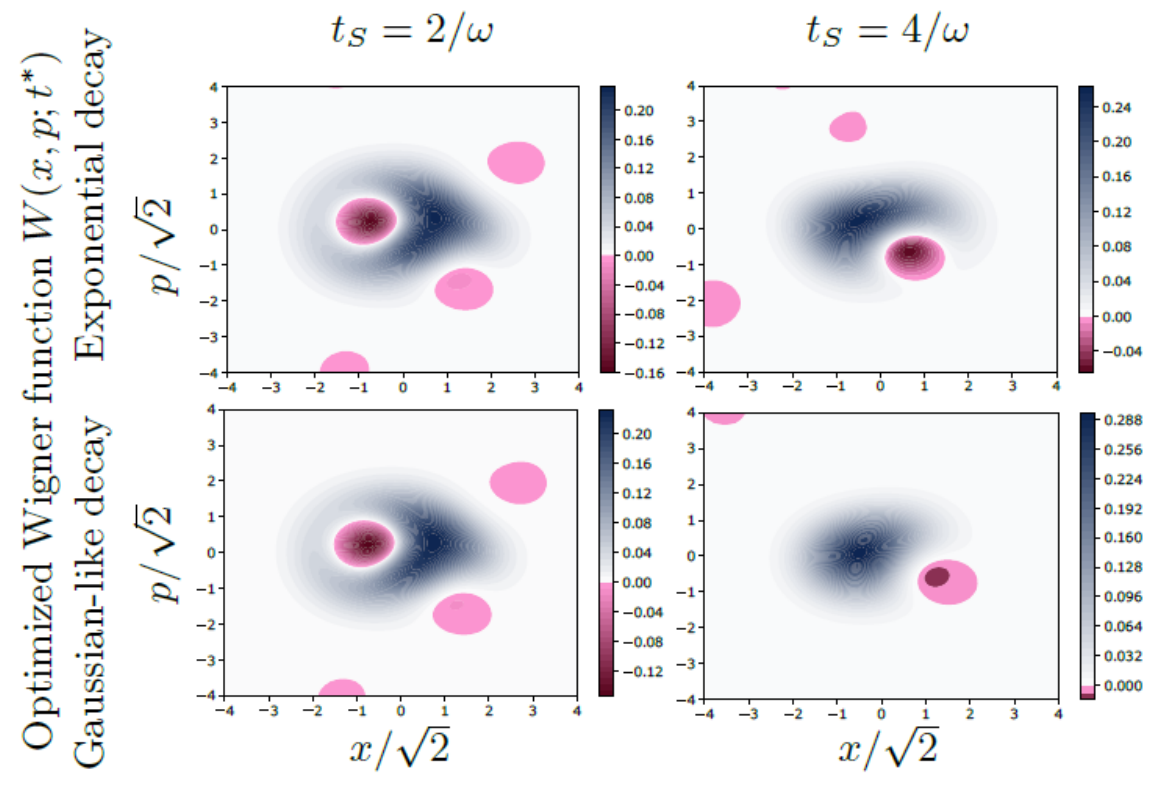}
\caption{Negativity $N(t^*)$ optimized Wigner functions for various decoupling time profiles and timescales of the unstable interaction $g_A(t)$, see Eqs.~\eqref{eq-H-Rabi-zxdriven-direct},~\eqref{eq-decay-profiles}, respectively. The parameters used are the same as in the previous plot: $\omega_0=2$, $\omega=1$, $\omega_A=2.4$, $g_R=0.6$, $g_{A}(0)=0.8$, $T=2\cdot 10^{-2}$, and $\kappa=2\cdot 10^{-3}$.}
\label{fig-negativity-tS-exp-gauss}
\end{figure}

For a more complete information, the resulting optimal Wigner functions for different decoupling profiles are shown in Fig.~\ref{fig-negativity-tS-exp-gauss}. The upper row represents the exponential decay, whereas the lower row the Gaussian-like decay profile of the unstable interaction $g_A(t)$, see Eqs.~\eqref{eq-H-Rabi-zxdriven-direct},~\eqref{eq-decay-profiles}, and Figs.~\ref{fig1},~\ref{fig-negativity-tS-exp-gauss}, respectively. For these particular examples of decay profiles, the Wigner functions do not differ substantially up to decay times $t_S\lesssim 2/\omega$, as shown in the left column. On contrary, for decay times $t_S\gtrsim 4/\omega$, right column, the decay profile does influence the Wigner function considerably, as the negativity vanishes faster for the Gaussian profile case. 

\section{Further TLS number scaling}
\label{sec:s-scaling}
\begin{figure}[b!t]
\includegraphics[width=.8\columnwidth]{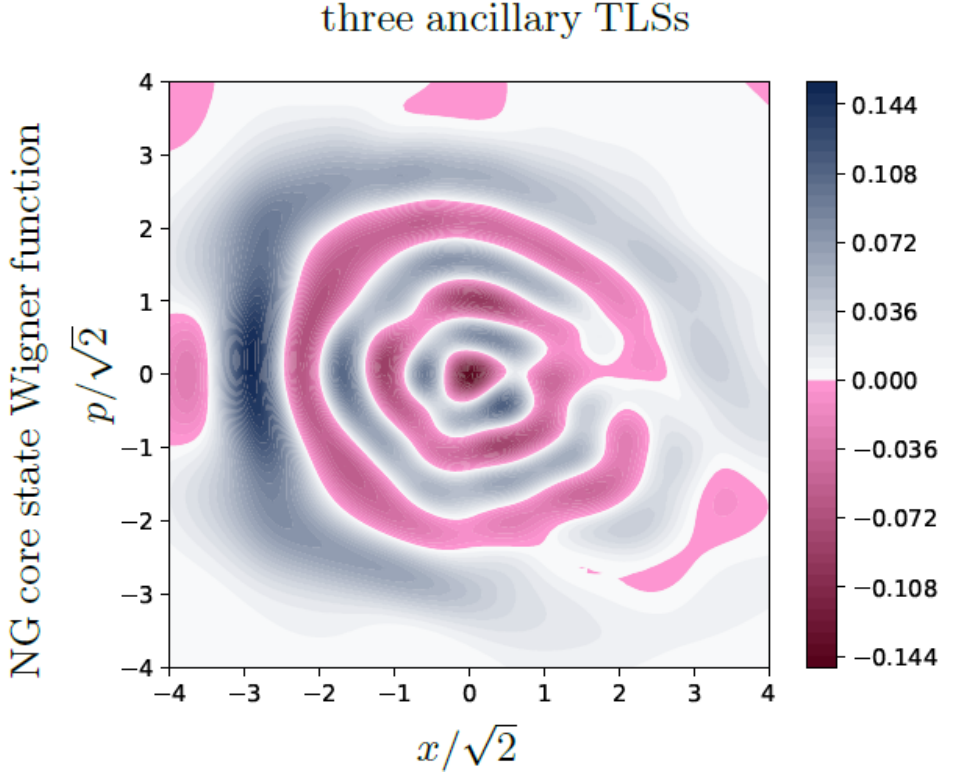}
\caption{The NG core state of LHO, if three ancillary TLSs are used during the initial thermalization stage. The dominant contribution stems from the Fock state $\ket{n}$ with $n=5$, see Fig.~\ref{fig-matrix-3A}. All ancillary TLSs were assumed to have the same frequency $\omega_A$, whereas the parameters used in simulation are the same as in Fig.~\ref{fig2}.} 
\label{fig-wigner-3A}
\end{figure}

\begin{figure}[ht]
\vspace{-.3cm}
\includegraphics[width=.7\columnwidth]{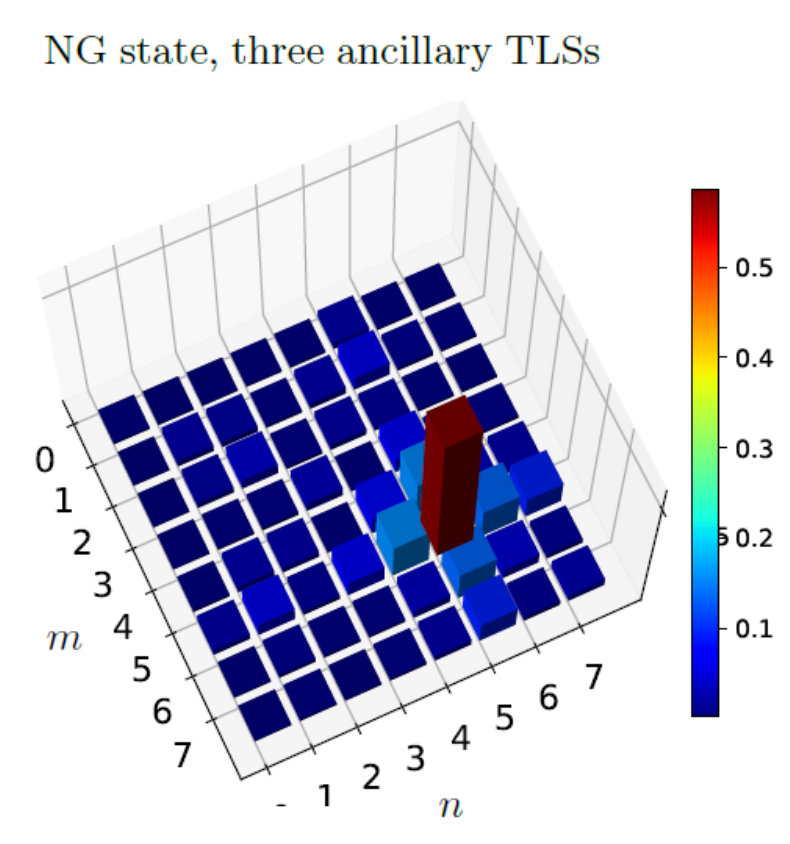}
\caption{Modules $|\rho_{mn}|$ of the density matrix elements of NG core state $\op{\rho}_3^{NG}=\op{S}(s_3)\op{D}(\beta_3)\op{\rho}_3^*\op{D}(-\beta_3)\op{S}^\dag(s_3)$ for three ancillary TLSs, see~Fig.~\ref{fig-wigner-3A}, with optimized parameters $s_3$, $\beta_3$, obtained numerically as described in Sec.~\ref{sec-coherence}. }
\label{fig-matrix-3A}
\end{figure}

This Appendix reveals the possibility of further ancillary TLSs number up-scaling, namely to three TLSs. Such protocol yields NG core states, see Sec.~\ref{sec-coherence}, with more complex structure determined by the higher Fock states dominating the NG core state, in particular $n=5$ with coherent contributions of states $n^\prime =n\pm 1$, as presented in Figs.~\ref{fig-wigner-3A}~and~\ref{fig-matrix-3A}. On the other hand, obtaining such higher photon-number states is counter-weighted by increased system complexity and non-negligibly higher need of computational resources in case of numerical simulations. 





\end{document}